\documentclass[pra,twocolumn,superscriptaddress,showpacs,longbibliography]{revtex4-1}
\usepackage{amssymb}
\usepackage{amsmath}
\usepackage{amsfonts}
\usepackage{bm}
\usepackage{graphicx}
\usepackage[dvipsnames]{xcolor}
\usepackage{calc} % to do for example 0.5\columnwidth+0.2\columnsep
\usepackage{epsfig}
\usepackage{epstopdf}
\usepackage{natbib}
\usepackage{hyperref}
\begin{document}

\title{Circuit Quantum Electrodynamics Simulator of Flat Band Physics in Lieb lattice}

\author{Zi-He Yang}
\affiliation{Wuhan National Laboratory for Optoelectronics and School of Physics, Huazhong University of Science and Technology, Wuhan, 430074, China}

\author{Yan-Pu Wang}
\affiliation{Wuhan National Laboratory for Optoelectronics and School of Physics, Huazhong University of Science and Technology, Wuhan, 430074, China}

\author{Zheng-Yuan Xue}
\affiliation{Guangdong Provincial Key Laboratory of Quantum Engineering and Quantum Materials, School of Physics and Telecommunication Engineering, South China Normal University, Guangzhou 510006, China}

\author{Wan-Li Yang}
\affiliation{State Key Laboratory of Magnetic Resonance and Atomic and Molecular Physics, Wuhan Institute of Physics and Mathematics, Chinese Academy of Sciences, Wuhan 430071, China}

\author{Yong Hu}
\email{huyong@mail.hust.edu.cn}
\affiliation{Wuhan National Laboratory for Optoelectronics and School of Physics, Huazhong University of Science and Technology, Wuhan, 430074, China}

\author{Jin-Hua Gao}
\email{jinhua@hust.edu.cn}
\affiliation{Wuhan National Laboratory for Optoelectronics and School of Physics, Huazhong University of Science and Technology, Wuhan, 430074, China}

\author{Ying Wu}
\affiliation{Wuhan National Laboratory for Optoelectronics and School of Physics, Huazhong University of Science and Technology, Wuhan, 430074, China}

\begin{abstract}
The concept of flat band plays an important role in strongly-correlated many-body physics. However, the demonstration of the flat band physics is highly nontrivial due to intrinsic limitations in conventional condensed matter materials. Here we propose a circuit quantum electrodynamics simulator of the 2D Lieb lattice exhibiting a flat middle band. By exploiting the parametric conversion method, we design a photonic Lieb lattice with \textit{in situ} tunable hopping strengths in a 2D array of coupled superconducting transmissionline resonators.
Moreover, the flexibility of our proposal enables the incorporation of both the artificial gauge field and the strong photon-photon interaction in a time- and site-resolved manner. To unambiguously demonstrate the synthesized flat band, we further investigate the observation of the flat band localization of microwave photons through the pumping and the steady-state measurements of only few sites on the lattice.
Requiring only current level of technique and being robust against imperfections in realistic circuits, our scheme can be readily tested in experiment and may pave a new way towards the realization of exotic photonic quantum Hall fluids including anomalous quantum Hall effect and bosonic fractional quantum Hall effect without magnetic field.
\end{abstract}


%42.50.Pq	Cavity quantum electrodynamics; micromasers
%85.25.Cp	Josephson devices
%42.25.Hz	Interference
%63.20.Pw	Localized modes

\maketitle

\section{Introduction}
\label{Sec Intro}

Quantum particles in a periodical crystal normally move freely except for a renormalized effective mass defined by the band dispersion. However, there exist structures exhibiting a completely flat band (FB) with an infinitely large effective mass \cite{SondhiFB2013,BergholtzFB2013}. As the kinetic energy is completely quenched, the interaction becomes dominant, making the FB an ideal platform of investigating strongly-correlated many-body physics including ferromagnetism, Wigner crystals, and
fractional quantum Hall states in the absence of magnetic field \cite{DasSarmaFB2007PRL, WuCJFB2010PRA,BernevigFB2011PRX,ShengDNFB2011NC,BerciouxPRA2011,BerciouxPRB2011,MarderPRA2016}. Nevertheless, despite the extensive efforts in the past decades, the proposed FB physics can still hardly be achieved in conventional electronic systems due to realistic reasons, e.g. the constraints of materials, the lack of controllability, and the co-existing complicated mechanisms.

Meanwhile, in recent years there have been ideas emerged that the similar lattice configurations can be built in controllable artificial photonic metamaterials \cite{PhotonSimulatorReviewNP2012,CarusottoQFLRMP2013,MukherjeeLiebFBPRL2015,VicencioLiebFBPRL2015,PolaritonFBPRl2015,FB1Darxiv2015,FBCondensePRL2016}. Compared with their electronic counterparts, these photonic simulators provide not only the same FB but also totally different non-equilibrium, charge-neutral, and bosonic properties, leading to both new physics and new challenges in theories and experiments. Motivated by these advances, in this manuscript we propose a circuit quantum dynamics (QED) simulator \cite{JQYouReview,DevoretReview2013,KochReview1,KochReview2} of the Lieb lattice, which is one of the most celebrated and important 2D lattices with FB configuration \cite{WuYSFB2014}. Here the lattice is constructed by superconducting transmissionline resonators (TLRs) coupled with superconducting quantum inteference devices (SQUIDs) \cite{FelicettiPRL2014,WangYPChiral2015,WangYPNPJQI2015}, and the electrons are substituted by the microwave photons. While the photonic Lieb lattice has already been realized recently in the context of photonic crystal with the FB localization of non-interacting photons been observed \cite{MukherjeeLiebFBPRL2015,VicencioLiebFBPRL2015}, our proposal takes the advantages of flexibility and tunability of superconducting quantum circuit (SQC). The first distinct merit of our scheme is that we synthesize the photon hopping by the parametric frequency conversion (PFC) approach, which is relatively simple in experimental setup and feasible with current technology \cite{WangYPChiral2015,WangYPNPJQI2015,NISTParametricConversionNP2011,NISTHongOuMandelPRL2012}. This PFC method can lead to the unprecedented \textit{in situ} tunable hopping strength, making the introduction of synthetic gauge fields for the neutral photons possible. In addition, the demonstrated strong coupling between superconducting qubits and TLRs \cite{JQYouReview,DevoretReview2013} allows the immediate incorporation of effective photon-photon interactions, which is crucial for strongly-correlated physics and has already attracted attentions in the very latest research of FB in 1D \cite{PolaritonFBPRl2015,FB1Darxiv2015,FBCondensePRL2016}. We further study the observation of localization-in-continuum modes in this architecture, which serves as the unambiguous evidence of the synthetic FB feature. Our discussions and numerical simulations based on realistic parameters pinpoint that a rather flat band structure can be obtained even in the presence of the various imperfection factors, and the proposed FB localization can consequently be observed through the steady-state photon number (SSPN) detection of only few sites on the lattice.

%This manuscript is organized as follows. We describe in Sec. \ref{Sec cQEDLieb} the circuit QED implementation of the Lieb lattice and then discuss in detail the influence on the FB from potential imperfections in Sec. \ref{Sec NNNandFE}. The probing of the FB localized modes is analyzed in Sec. \ref{Sec LIC}, and several generalizations of our scheme are discussed in Sec. \ref{Sec Discussion}. The conclusion of our paper is made in Sec.\ref{Sec Conclusion}.

\section{The circuit QED Lieb lattice}
\label{Sec cQEDLieb}

\begin{figure}[tbhp!]
\begin{center}
\includegraphics[width=0.48\textwidth]{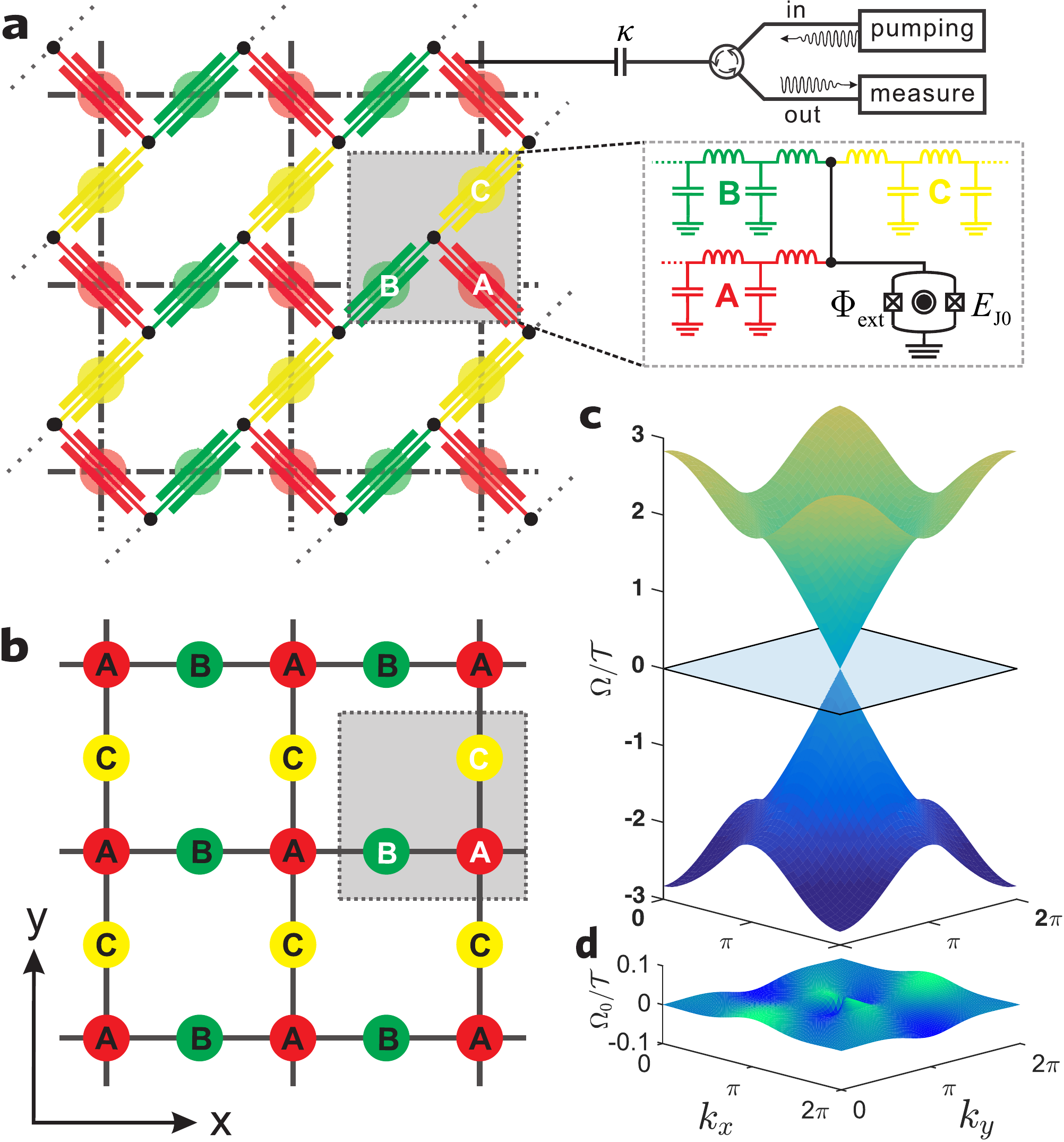}
\end{center}
\caption{\label{Fig Lattice}(Color Online) (a) Circuit QED implementation of the Lieb lattice. The TLRs are exploited as the photonic lattice sites and the grounding SQUIDs induce the coupling between them. The colors of the TLRs label their different lengths and consequently different eigenfrequencies. (b) Sketch of the Lieb lattice composed of unit-cells with three sites labeled A, B, and C, and the $A \Leftrightarrow B$ and $A \Leftrightarrow C$ bonds along the horizontal and vertical directions, respectively. (c) Band structure of the Lieb lattice in the first Brillouin zone. In the ideal situation a non-dispersive FB emerges as the middle band of the lattice. (d) Configuration of the middle band in the presence of the NNN couplings and with the application of the interlaced unit-cell strategy.}
\end{figure}

Here we propose the circuit QED lattice shown in Fig.~\ref{Fig Lattice}(a) as the SQC realization of the Lieb lattice, which consists of three types of TLRs differed by their lengths and placed in an interlaced bricklayer form. These TLRs play the corresponding roles of the A, B, and C sites of the line-centered-square lattice depicted in Fig.~\ref{Fig Lattice}(b). At their ends, the TLRs are grounded by SQUIDs with effective inductances much smaller than those of the TLRs. Due to their very small inductances, the grounding SQUIDs impose the low-voltage shortcut boundary conditions for the TLRs \cite{FelicettiPRL2014,WangYPChiral2015,WangYPNPJQI2015}. The lowest eigenmodes of the lattice can then be approximated by the individual $\lambda/2$ modes of the TLRs, and the lattice can be described by the Hamiltonian
\begin{equation}
\mathcal{H}_{\mathrm{S}}= \sum_{\mathbf{r}} \omega_{\mathrm{A}} A_{\mathbf{r}}^{\dagger} A_{\mathbf{r}}+\omega_{\mathrm{B}} B_{\mathbf{r}}^{\dagger} B_{\mathbf{r}}+\omega_{\mathrm{C}} C_{\mathbf{r}}^{\dagger} C_{\mathbf{r}},
\label{Eqn Hami}
\end{equation}
where $\alpha_{\mathbf{r}}^{\dagger}$/$\alpha_{\mathbf{r}}$ are the creation/annihilation operators of the $\alpha$th site in the $\mathbf{r}$th unit-cell for $\alpha=\mathrm{A,B,C}$, and $\omega_{\alpha}$ are their eigenfrequencies. Hereafter we specify $(\omega_{\mathrm {A}},\omega_{\mathrm {B}},\omega_{\mathrm {C}})=(\omega_{\mathrm {0}},\omega_{\mathrm{0}}-\Delta, \omega_{\mathrm{0}}+2\Delta)$ with $\omega_{\mathrm{0}}/2\pi \in \left[10,15\right]\,\mathrm{GHz}$ and $\Delta/2\pi \in \left[1,2\right]\,\mathrm{GHz}$. Such configuration is for the following application of the PFC method and can be experimentally realized through the length selection of the TLRs in the millimeter range \cite{NISTParametricConversionNP2011,NISTHongOuMandelPRL2012,NISTParametricCouplingPRL2014,NISTCoherentStateAPL2015}. We refer to Appendix. \ref{App Eigen} for detailed characterization of the eigenmodes and the estimation of the circuit parameters.

We further consider the implementation of the effective photon hopping on the lattice, taking the general form
\begin{align}
\label{Eqn Lieb2}
  \mathcal{H}_{\mathrm{L}}=\sum_{\langle (\mathbf{r},\alpha), (\mathbf{r^\prime},\beta)\rangle}
  \mathcal{T}_{\mathbf{r},\alpha}^{\mathbf{r}^{\prime},\beta}\beta_{\mathbf{r}^\prime}^{\dagger}
  \alpha_{\mathbf{r}}e^{i\theta_{\mathbf{r},\alpha}^{\mathbf{r}^{\prime},\beta}},
\end{align}
in the rotating frame of $\mathcal{H}_{\mathrm{S}}$. Here $\mathcal{T}_{\mathbf{r},\alpha}^{\mathbf{r}^{\prime},\beta}$
is the real $\left( \mathbf{r},\alpha \right) \rightarrow \left( \mathbf{r^{\prime}},\beta \right)$ hopping strength and $\theta_{\mathbf{r},\alpha}^{\mathbf{r}^{\prime},\beta}=\int_{\mathbf{r},\alpha}^{\mathbf{r}^\prime,\beta} \mathbf{A(x)}\cdot\mathrm{d}\mathbf{x}$
is the $\left( \mathbf{r},\alpha \right) \rightarrow \left( \mathbf{r^{\prime}},\beta \right)$ hopping phase manifesting the presence of an Abelian gauge potential $\mathbf{A(x)}$ \cite{BernevigBook}. For each plaquette of the lattice, the loop summation of the hopping phases can be regarded as the synthesized magnetic field for the microwave photons, i. e. $\oint \mathbf{A(x)}\cdot\mathrm{d}\mathbf{x}=\iint \mathbf{B(x)}\cdot\mathrm{d}\mathbf{S}$. Meanwhile, as the physical coupling between two TLRs usually takes real coupling constants \cite{UnderwoodDisorderPRA2012,LiebSQUIDCouplingNJP2012}, we exploit the alternative dynamic modulation method to implement the general complex $\mathcal{H}_{\mathrm{L}}$ \cite{FanSHDMNP2012,NISTParametricConversionNP2011,NISTHongOuMandelPRL2012,NISTHongOuMandelPRL2012,DCEexperimentNature2011}. The grounding SQUIDs can be regarded as tunable inductances which can be a.c. modulated by external magnetic flux oscillating at very high frequencies \cite{DCEexperimentNature2011}. Such a.c. modulation introduces a small fraction
\begin{equation}
 \mathcal{H}_\mathrm{a.c.}=\sum_{\langle (\mathbf{r},\alpha) , (\mathbf{r^{\prime}},\beta) \rangle} \mathcal{T}^{\mathrm{ac}}_{(\mathbf{r},\alpha), (\mathbf{r^{\prime}},\beta)}(t)(\alpha_{\mathbf{r}}+\alpha_{\mathbf{r}}^{\dagger}) (\beta_{\mathbf{r^{\prime}}}+\beta_{\mathbf{r^{\prime}}}^{\dagger}),
\end{equation}
 in addition to the d.c. contribution of the grounding SQUIDs which proves to be irrelevant because the neighboring TLRs are largely off resonant (see Appendix. \ref{App Eigen}). We then assume that the a.c. modulation of the grounding SQUIDs contain two tones with frequencies $\Delta$ and $2\Delta$, which induce the horizontal $A \Leftrightarrow B$ and vertical $A \Leftrightarrow C$ PFC bonds by bridging their frequency gaps respectively. When experiencing this PFC process, the microwave photons will adopt the phases of the a.c. modulating pulses, leading to the effective controllable complex hopping constants \cite{WangYPChiral2015,WangYPNPJQI2015,FanSHDMNP2012}. Moreover, from Fig. \ref{Fig Lattice}(a) it can be figured out that each of the vertical and horizontal hopping bonds can be independently controlled by a modulating tone threaded in one of the grounding SQUIDs, leading to the site-resolved control of both the hopping strengths $\mathcal{T}_{\mathbf{r},\alpha}^{\mathbf{r}^{\prime},\beta}$ and the hopping phases $\theta_{\mathbf{r},\alpha}^{\mathbf{r}^{\prime},\beta}$. Our further estimations show that the hopping strengths can be designed in the range $\mathcal{T}_{\mathbf{r},\alpha}^{\mathbf{r}^{\prime},\beta}/2\pi \in \left[ 5, 15 \right]\, \mathrm{MHz}$. The derivation of the described dynamic modulation method is detailed in Appendix. \ref{App Eigen}, and a set of typical parameters is proposed in Tab. \ref{Tab para}, which is selected based on recent experiments of parametric processes in SQC and will be used for the numerical simulations throughout this paper.

\begin{table}[tbh!]
\centering
\begin{tabular}{p{0.24\textwidth}p{0.24\textwidth}}
  \hline
  TLRs parameters &  \\
  \hline
  unit inductance/capacitance & $l=4.1\times 10^{-7}\,\mathrm{H}\cdot \mathrm{m}^{-1}$, $c=1.6\times 10^{-10}\,\mathrm{F}\cdot \mathrm{m}^{-1}$ \cite{NISTParametricConversionNP2011,NISTHongOuMandelPRL2012,NISTParametricCouplingPRL2014}\\
  lengths of the TLRs &  $L_{\mathrm{A}}=5.6\,\mathrm{mm}$, $L_{\mathrm{B}}=6.8\,\mathrm{mm}$, $L_{\mathrm{C}}=4.1\,\mathrm{mm}$ \cite{DCEexperimentNature2011,NISTParametricConversionNP2011,NISTHongOuMandelPRL2012}\\
\\
  \hline
  SQUIDs &\\
  \hline
  maximal critical currents & $I_{\mathrm{J0}}=75.5\,\mu \mathrm{A}$ \cite{DCEexperimentNature2011,NISTParametricConversionNP2011,YuYangScience2002,MartinisPhaseQubitPRL2002}\\
  d.c. flux bias points & $\Phi_\mathrm{ex}^\mathrm{dc}=0.37\Phi_0$ \cite{NISTParametricConversionNP2011,NISTHongOuMandelPRL2012}\\
  effective critical currents & $I_{\mathrm{J}}=30\,\mu \mathrm{A}$\\
  junction capacitances & $C_\mathrm{J}=0.5$ $\mathrm{pF}$ \cite{YuYangScience2002,MartinisPhaseQubitPRL2002}\\
  a.c. modulation amplitudes & $\Phi_\mathrm{CA}=1.3\%\Phi_0$, $\Phi_\mathrm{BA}=0.9\%\Phi_0$ \cite{NISTParametricConversionNP2011}\\
  \hline
  Eigenmodes \& coupling & \\
  \hline
  eigenfrequencies & $\omega_\mathrm{A}/2\pi=11\,\mathrm{GHz}$, $\omega_\mathrm{B}/2\pi=9\,\mathrm{GHz}$, $\omega_\mathrm{C}/2\pi=15\,\mathrm{GHz}$ \cite{DCEexperimentNature2011,NISTParametricConversionNP2011,NISTHongOuMandelPRL2012}\\
  uniform decay rate & $\kappa/2\pi=100\,\mathrm{kHz}$ \cite{DCEexperimentNature2011,NISTParametricConversionNP2011,YuYangScience2002,MartinisPhaseQubitPRL2002,NISTCoherentStateAPL2015}\\
  hopping constant & $\mathcal{T}_{\mathbf{r},\alpha}^{\mathbf{r}^{\prime},\beta}/2\pi=\mathcal{T}/2\pi=10$ $\mathrm{MHz}$\\
  \hline
\end{tabular}
\caption{Representative parameters of the proposed circuit selected based on recently-reported experiments.}
\label{Tab para}
\end{table}

In addition, we should mention that the proposed lattice is not limited by the Lieb lattice configuration focused in this manuscript. By adding an additional $3\Delta$ tone in each of the grounding SQUIDs, we can straightforwardly get a stretched Kagom\'e lattice by opening the $B \Leftrightarrow C$ hopping bonds (Fig. \ref{Fig Lattice}(a)). This generalization is natural in the sense that the Kagom\'e lattice and the honeycomb lattice (i.e. the stretched bricklayer in Fig. \ref{Fig Lattice}(a)) are the line-graphs of each other (i.e. the roles of bonds and sites are exchanged) \cite{WuYSFB2014}. This facility may pave an alternative way of investigating anomalous quantum Hall effect and topologically nontrivial FB in the future \cite{BerciouxPRA2011,BerciouxPRB2011,PetrescuAQHEPRA2012,MarderPRA2016}.

Before proceeding, we offer a brief remark about the independent addressing of the grounding SQUIDs on the lattice. In the past decade, the individual flux control has already been achieved in coupled superconducting flux qubits \cite{PlantenbergCNOTNature2006,PloegCoupleFQ2007}, where several coils have been applied to manipulate the d.c. and a.c. flux biases threaded in neighboring flux qubits with both the loop sizes of the qubits and the distance between the qubits being at the range of micronmeters. On the other hand, the spacing between the grounding SQUIDs in our proposal is at the level of the length scale of the TLRs ($\sim$ millimeters, see Tab. \ref{Tab para}), which is by several orders larger than the distances between the loops of the flux qubits. From this point of view, the requirement of individual flux addressing is more weak than those of the reported experiments, because larger distance between the considered SQUIDs indicates smaller cross-talk and easier fabrication of the biasing coils. When the scaled-up lattice is taken into consideration, the requirement of controlling many SQUIDs individually leads to more complicated coil setup than that of the few-qubit case. Meanwhile, the very large spacing between the grounding SQUIDs still offers enough room of design. One potential solution is that we add an additional layer of antenna on top the sample that contains the array of the TLRs. Here we should notice that increasing research interest has recently been attracted by the design of scalable architecture that combines various quantum elements into a complex device without compromising their performance, and a multilayer microwave integrated quantum circuit platform has already been developed to couple a large number of circuit components through controllable channels while suppressing any other interactions \cite{YaleMultilayerNPJQI2016}. Therefore, it is our opinion that the requirement of individual addressing does not place a hindrance towards the experimental realization of the proposed scheme.

\section{The flatness of the synthesized middle band}
\label{Sec NNNandFE}

\begin{figure}[tbh!]
\centering
\includegraphics[width=0.48\textwidth]{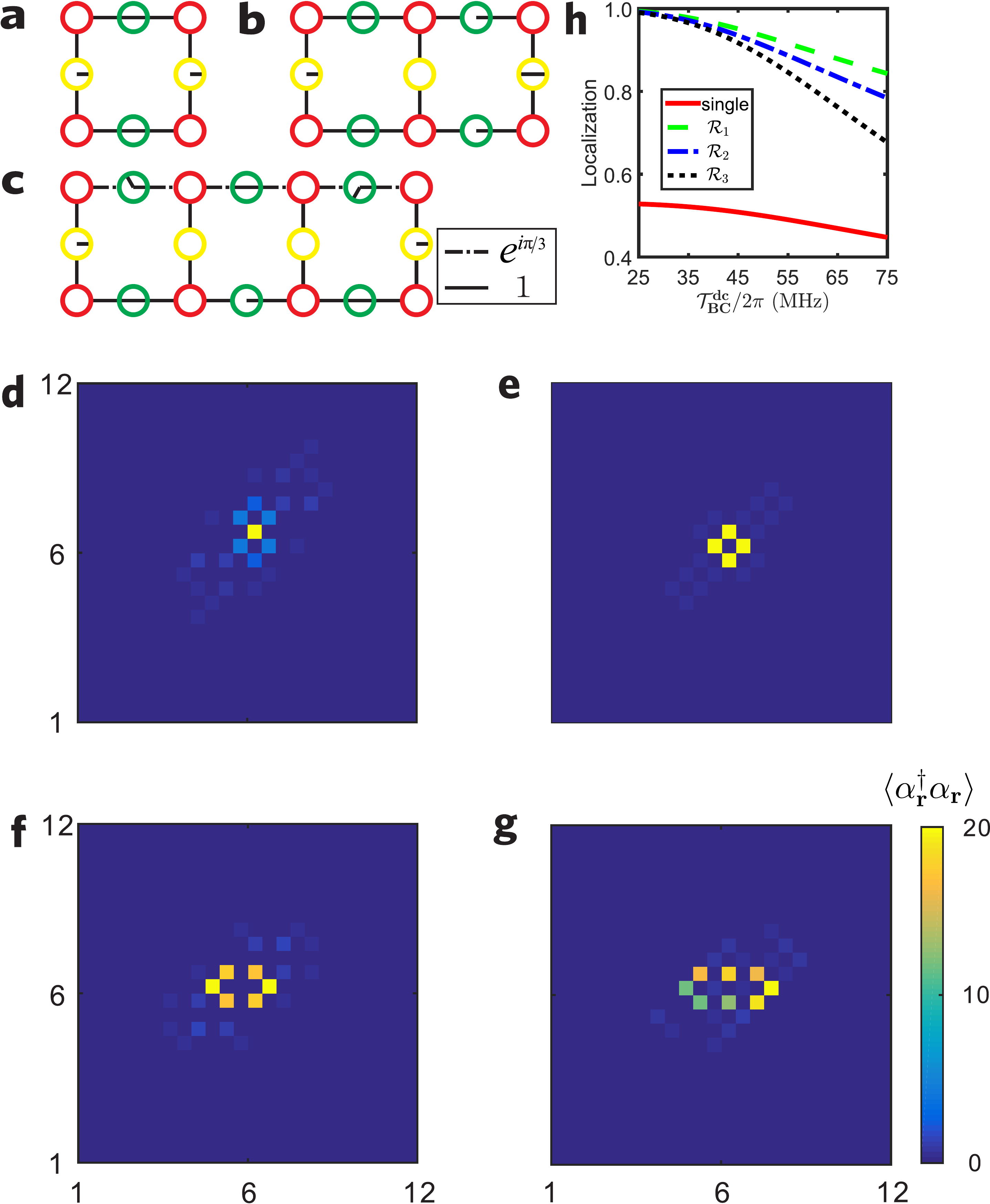}
\caption{\label{Fig interference}(Color Online) RM states and RM pumping of the proposed lattice.
The RM states fulfilling Eq.~(\ref{Eqn DIcondition}) usually have uniform amplitudes on the B and C sites but zero on the A sites.
The relative phases between the sites are denoted by the angles in the rounds.
(a) and (b) represent two RM states in the absence of magnetic field $| \Phi_{1} \rangle =\mathcal{R}_{1} | 0 \rangle=2^{-1}(C_{x_0,y_0}^{\dagger}-B_{x_0+1,y_0}^{\dagger}+C_{x_0+1,y_0}^{\dagger}-B_{x_0+1,y_0+1}^{\dagger})| 0 \rangle$ and $|\Phi_{2} \rangle = \mathcal{R}_{2} | 0 \rangle=6^{-1/2}(C_{x_0,y_0}^{\dagger}-C_{x_0+2,y_0}^{\dagger}-B_{x_0+1,y_0}^{\dagger}-B_{x_0+1,y_0+1}^{\dagger}+B_{x_0+2,y_0}^{\dagger}+B_{x_0+2,y_0+1}^{\dagger})| 0 \rangle$, with $\mathbf{r}_0=[x_0,y_0]$ being a particular unit-cell, and (c) represents a RM state $| \Phi_{3} \rangle =\mathcal{R}_{3} | 0 \rangle$ in the presence of synthetic magnetic field $\mathbf{B}=\mathbf{e}_z 2\pi/3$. The solid and dashed lines correspond to the bonds with zero and $\pi/3$ hopping phases, respectively.
(d)--(h) depict the FB localization of a lattice under pumping, with the NNN coupling in Eq. (\ref{Eqn NNND}) taken into account. (d) corresponds to the SSPN distribution of the single-pumping situation, while (e)--(g) correspond to the RM pumping in (a)---(c), respectively. The FB localization is further quantified by the localization factor versus $\mathcal{T}_{BC}^\mathrm{dc}$ shown in (h).
}
\end{figure}

The novelty of the Lieb lattice lies in the existence of the FB configuration. To be concrete let us focus on the specific situation of Eq. (\ref{Eqn Lieb2}) with $\mathcal{T}_{\mathbf{r},\alpha}^{\mathbf{r}^{\prime},\beta}=\mathcal{T}$ and $\mathbf{A(x)}=0$, in which the band structure of the lattice becomes
\begin{align}
\label{Eqn flateigen}
\Omega_{\pm}=\pm 2 \mathcal{T} \sqrt{\cos^2 \frac{k_x}{2}+\cos^2 \frac{k_y}{2}},\quad \Omega_{0}=0,
\end{align}
with $k_x,k_y \in [0,2\pi]$ being the pseudo-momentums (detailed band structure calculation method can be found in Appendix. \ref{App Band}, see also Ref. \cite{BergholtzFB2013}). As shown in Fig. \ref{Fig Lattice}(c), Eq. (\ref{Eqn flateigen}) provides a non-dispersive middle band with zero eigenvalue and a Dirac cone structure around the three-band touching point $(k_{x},k_{y})=(\pi,\pi)$. The existence of the FB implies the emergence of exotic FB localization on the perfect periodic lattice, which is distinct from localization induced by disorder and should be interpreted by the mechanism of destructive interference \cite{WuYSFB2014}: A single-particle state $| \Phi \rangle =\sum_{\mathbf{r}, \alpha}\,\mathcal{P}_{\mathbf{r}, \alpha}\alpha_{\mathbf{r}}^{\dagger}| 0 \rangle $ satisfies $\mathcal{H}_{\mathrm{L}} | \Phi \rangle=0$ iff the condition
\begin{align}
\label{Eqn DIcondition}
\sum_{\langle \left(\mathbf{r},\alpha\right),\left(\mathbf{r^{\prime}},\beta\right) \rangle}\, \mathcal{P}_{\mathbf{r^\prime}\beta}=0, \qquad \forall \quad \mathbf{r},\alpha,
\end{align}
is met. The ring mode (RM) state  $| \Phi_{1} \rangle $ shown in Fig.~\ref{Fig interference}(a) is a representative example satisfying Eq.~(\ref{Eqn DIcondition}): If the photon wants to run away from this plaquette, it has to pass first through the four A sites on the corner. However, Eq.~(\ref{Eqn DIcondition}) guarantees the coherent cancelation of the photon flows towards the A sites, leaving $| \Phi_{1} \rangle $ localized. The similar situation is also valid for another RM state $|\Phi_{2} \rangle $ shown in Fig. \ref{Fig interference}(b), which contains two plaquette and can be regarded as the superposition of two single-plaquette RM states in Fig. \ref{Fig interference}(a). A direct generalization of this observation figures out that the number of independent states fulfilling Eq. (\ref{Eqn DIcondition}) equals the number of independent plaquette on the lattice, and it is this set of states spanning the middle FB in Fig. \ref{Fig Lattice}(c).

\begin{figure}[tbh!]
\centering
\includegraphics[width=0.48\textwidth]{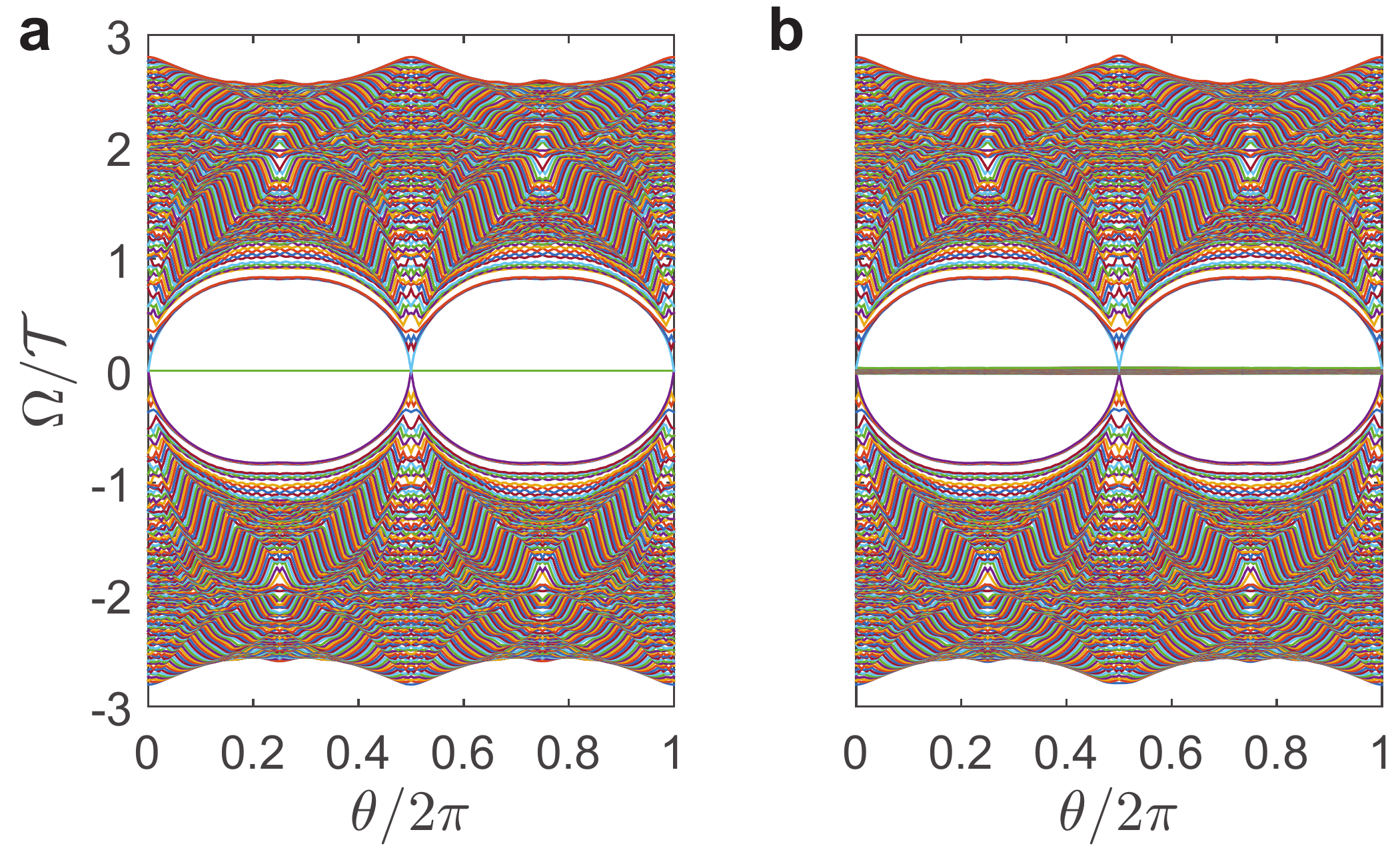}
\caption{\label{Fig Butterfly}(Color Online) Hofstadter butterfly spectrum of the proposed Lieb lattice. Here we consider a lattice consisting of $12\times 12$ unit-cells with open boundary condition. (a) corresponds to the ideal situation where the NNN couplings do not exist, while (b) corresponds to the realistic case where the NNN channels shown in Eq. (\ref{Eqn NNND}) are taken into account. The eigenenergy spectrum is obtained by first writing down the $\theta$--dependent matrix $\mathcal{B}$ (its definition can be found in Sec. \ref{Sec LIC}) and  then diagonalizing it in a brutal-force way.}
\end{figure}

Moreover, we can go beyond the non-magnetic situation $\mathbf{A(x)}=0$ to consider the synthetic FB in the presence of artificial gauge field, as the synthesized magnetic field for the charge-neutral photons can be introduced by exploiting the freedom of controllable hopping phase. Without loss of generality we choose Landau gauge that the vertical hopping phases are zero and the horizontal hopping phases are $\theta_{\mathbf{r},\alpha}^{\mathbf{r^\prime},\beta}=\int_{\mathbf{r},\alpha}^{\mathbf{r^\prime},\beta}\,\mathbf{A} \cdot \mathrm{d}\mathbf{r}$ for $\mathbf{r}=[m,n]$ with $\mathbf{A}=-\theta n\mathbf{e_x}$. The presence of the artificial magnetic field $\mathbf{B}=2\theta\mathbf{e_z}$ results in the fractral Hofstadter butterfly spectrum \cite{HofstadterButterflyPRB1976} with the middle band remained flat during the variation of $\theta$, as shown in Fig.~\ref{Fig Butterfly}(a). The preservation of the flatness can be illustrated by a specific example $\theta=\pi/3$ where the unit-cell of the lattice is enlarged by three times and the typical RM state takes the form shown in Fig.~\ref{Fig interference}(c). Just as the same as the previous non-magnetic situations in Figs. \ref{Fig interference}(a) and \ref{Fig interference}(b), each plaquette still corresponds to an independent RM state, and the number of independent states satisfying Eq. (\ref{Eqn DIcondition}) is unchanged in the presence of $\mathbf{B}$.

However, in realistic experiments there exists unavoidably imperfection factors breaking the ideal flatness of the middle band, including the residual d.c. mixing between TLRs, the fabrication errors of the circuit, and the background low-frequency noises. Understanding their effects is thus very important for our scheme. In what follows we discuss these imperfections with results showing that the induced effects are all much lower than the hopping strength $\mathcal{T}$, and some of them can be further suppressed through the slight refinement of the developed PFC method.

\textit{The background d.c. mixing.}---The background d.c. mixing between the physically neighboring TLRs can be characterized by the d.c. coupling strength $\mathcal{T}^\mathrm{dc}_{\alpha\beta}/2\pi \in [45,60] $ $\mathrm{MHz}$ (see Appendix. \ref{App Eigen}) and can result in Stark shifts of the TLRs and coupling between next-nearest-neighbor (NNN) TLRs. These effects can be understood by the dispersive coupling mechanism \cite{ZhengSBDispersivePRL2000}. Let us imagine a photon initially populated in a particular site $A_{m,n}$. It can hop to its neighbor $B_{m,n}$ via the d.c. coupling channel induced by their common grounding SQUID. Due to the large detuing between these two sites, the photon can not be stable in $B_{m,n}$, and its fate is either hopping back to $A_{m,n}$, resulting in a Stark shift ${\mathcal{T}^\mathrm{dc}_\mathrm{AB}}^2 /\Delta(A_{m,n}^{\dagger}A_{m,n}-B_{m,n}^{\dagger}B_{m,n}) $, or hopping further to $A_{m,n+1}$, resulting in an NNN coupling ${\mathcal{T}^\mathrm{dc}_\mathrm{AB}}^2/\Delta (A_{m+1,n}^{\dagger}A_{m,n}+\mathrm{h.c.}) $.
The Stark shifts and the NNN coupling then lead to a finite width $\sim {\mathcal{T}^\mathrm{dc}_{\alpha\beta}}^2/\Delta$ of the central band.

Meanwhile, these negative effects can be suppressed by the following methods: The Stark shifts can be cancelled by modifying the modulating frequencies of the grounding SQUIDs accordingly. For the NNN coupling, notice that the dispersive coupling mechanism relies on the frequency match of the NNN TLRs, i.e. the NNN hopping can effectively happen only if the NNN TLRs have the same eigenfrequencies \cite{ZhengSBDispersivePRL2000}. Therefore we exploit the interlaced unit-cell strategy where the eigenfrequencies of the modes in unit-cell $(m,n)$ are unchanged if $m+n$ is even and shifted up by $\Delta/3$ if $m+n$ is odd. With this configuration the NNN coupling between neighboring unit-cells are effectively suppressed and only the two ``diagonal'' NNN hopping channels need to be taken into consideration:
\begin{subequations}
\label{Eqn NNND}
\begin{alignat}{2}
C &\Leftrightarrow B \Leftrightarrow C,&\quad \frac{{\mathcal{T}_\mathrm{BC}^\mathrm{dc}}^2}{3\Delta}&\sum_{m,n} \, (C^{\dagger}_{m,n}C_{m+1,n+1}+\mathrm{h.c.}),\label{Subeq e}\\
B &\Leftrightarrow C \Leftrightarrow B,&-\frac{{\mathcal{T}_\mathrm{BC}^\mathrm{dc}}^2}{3\Delta}&\sum_{m,n} \, (B^{\dagger}_{m,n}B_{m+1,n+1}+\mathrm{h.c.}).\label{Subeq f}
\end{alignat}
\end{subequations}
Based on the proposed parameters in Tab. \ref{Tab para}, the strength of the residual NNN coupling can be estimated as
\begin{equation}
\frac{{\mathcal{T}_\mathrm{BC}^\mathrm{dc}}^2}{3\Delta} \approx 2\pi\times 0.6\,\mathrm{MHz} < 10^{-1}\mathcal{T}.
\end{equation}
The shapes of the middle band in this situation is calculated and plotted in Figs.~\ref{Fig Lattice}(d) and \ref{Fig Butterfly}(b) for the non-magnetic and magnetic situations, respectively. For the non-magnetic situation, we observe from Fig.~\ref{Fig Lattice}(d) that the flatness of the middle band is still preserved to some extent with bandwidth being  ${\mathcal{T}_\mathrm{BC}^\mathrm{dc}}^2/{2\Delta}$, while in the magnetic situation the degeneracy of the middle FB band is broken by the NNN coupling, indicated by the ``fat'' middle FB in Fig.~\ref{Fig Butterfly}(b) with the bandwidth similar to that of the non-magnetic case.

\textit{The fabrication error.}---The fabrication errors induce the deviations of the realized circuit parameters from the ideal settings (e.g. the lengths and the unit capacitances or inductances of the TLRs) and lead to the disorder $\delta\omega_{\mathbf{r},\alpha}$ of the eigenmodes' frequencies. Meanwhile, with developed microelectronic techniques such fabrication-induced disorder can be pushed to the level of $10^{-4}$ \cite{UnderwoodDisorderPRA2012}, which corresponds to $\delta\omega_{\mathbf{r},\alpha} \sim 10^{-1}\mathcal{T}$. Moreover, one can similarly cancel the fabrication-induced frequency shift by adjusting the frequencies of the two-tone PFC pulses in the grounding SQUIDs. With such refinement the fabrication-induced diagonal disorder can be effectively suppressed while the performance of the dynamic modulation method is not affected.

\textit{Low frequency $1/f$ noise.}---The low-frequency $1/f$ noise is ubiquitous in SQCs and its influence exceeds that of the thermodynamic noise \cite{FlickerRMP2014}. The $1/f$ noise in the proposed circuit can generally be traced back to the fluctuations of three degrees of freedom, namely the charge, the flux, and the critical current. Firstly, the proposed circuit is insensitive to the charge noise as it consists of only linear TLRs and grounding SQUIDs with very small anharmonicity. Such insensitivity roots in the same origin of the charge insensitivity of transmon qubits \cite{KochTransmonPRA2007}. Secondly, the flux $1/f$ noises penetrated in the loops of the grounding SQUIDs shift the d.c. bias of the grounding SQUIDs in a quasi-static way. The consequent effect is then the fluctuations
\begin{align}
\delta \omega_{\mathbf{r},\alpha} < 10^{-3} \mathcal{T},
\delta \mathcal{T}_{\mathbf{r},\alpha}^{\mathbf{r}^{\prime},\beta} < 10^{-4} \mathcal{T},
\end{align}
where the detailed evaluation is included in Appendix. \ref{App noise}. Both the diagonal and off-diagonal fluctuations are much smaller than the propose homogenous hopping strength $\mathcal{T}$. Therefore such fluctuations can influence negligibly on the flatness of the middle band. The effects of the critical current noise is similarly analyzed in Appendix. \ref{App noise}, with results indicating that the induced disorders are even smaller than those of the flux $1/f$ noises by several orders \cite{MartinisPhaseQubitPRL2002,MartinisFlickerPRL2007}. We thus come to the conclusion that our scheme is robust against the $1/f$ noise in SQC, leaving the residual NNN coupling the main and the only imperfection factor that should be taken care of.

\section{Probing the FB localization}
\label{Sec LIC}

The photonic nature of circuit QED allows the multiple occupation of a particular mode and the driving-dissipation competition. Such bosonic non-equilibrium feature can be exploited to demonstrate the FB localization on the proposed lattice, which serves as the undoubtful evidence and the quantification of the synthetic middle FB. Here we emphasize that the essential physics behind is the exotic middle FB of the Lieb lattice which leads to the novel steady states of the circuit. Explicitly, we study the coherent pumping described by $\mathcal{P}^{\dagger}\mathbf{a}e^{-i\Omega_{\mathrm{P}} t}+\mathrm{h.c.}$ in the rotating frame of $\mathcal{H}_{\mathrm{S}}$ where $\mathcal{P}$ and $\mathbf{a}$ are the vectors composed of the pumping strengths and the annihilation operators on the lattice, respectively, and $\Omega_{\mathrm{P}}$ is the monochromatic detuning. The steady state of the lattice is determined by
\begin{equation}
\label{Eqn SSnumber}
i\frac{\mathrm{d}{\langle \mathbf{a}  \rangle}}{\mathrm{d}t}
=\left[\mathcal{B} -\left(\Omega_{\mathrm{P}}+\frac{1}{2} i\kappa\right) \mathcal{I}\right]\langle \mathbf{a} \rangle + \mathcal{P}=0,
\end{equation}
where $\kappa$ is the assumed uniform decay rate of the TLRs and the matrix $\mathcal{B}$ is defined by $\mathbf{a}^{\dagger}\mathcal{B}\mathbf{a}=\mathcal{H}_{\mathrm{L}}$. We then consider the pumping and the consequent steady states of a lattice consisting of $12\times 12$ unit-cells with the open boundary condition imposed and the NNN coupling in Eq. (\ref{Eqn NNND}) taken into consideration. We study the following four pumping situations $\mathcal{P}^{\dagger}_{1}\mathbf{a}=\mathcal{T}_\mathrm{P}B_{x_0,y_0}$, $\mathcal{P}^{\dagger}_{2}\mathbf{a}=\mathcal{T}_\mathrm{P}\mathcal{R}_1$, $\mathcal{P}^{\dagger}_{3}\mathbf{a}=\mathcal{T}_\mathrm{P}\mathcal{R}_2$, and
$\mathcal{P}^{\dagger}_{4}\mathbf{a}=\mathcal{T}_\mathrm{P}\mathcal{R}_3$
as shown in Figs. \ref{Fig interference}(b)---(d), with $\mathbf{r}=[x_0,y_0]=[6,6]$, $\mathcal{T}_\mathrm{P}/2\pi=1\,\mathrm{MHz}$, and $\kappa/2\pi=100\,\mathrm{kHz}$ \cite{NISTParametricConversionNP2011,NISTHongOuMandelPRL2012,NISTCoherentStateAPL2015,DCEexperimentNature2011}.
The first three situations correspond to single-site or RM pumpings in the non-magnetic case, while the fourth corresponds to the RM pumping in the presence of $\mathbf{B}=\mathbf{e}_z 2\pi/3$.
The corresponding SSPN distributions are depicted in Figs.~\ref{Fig interference}(d)---\ref{Fig interference}(g), respectively. As shown in Fig.~\ref{Fig interference}(d), the steady state of the single-site pumping is extended around the unit-cells neighboring to the pumping site. Meanwhile, when the three RM pumpings are applied, the steady states become significantly localized, indicated by Figs.~\ref{Fig interference}(e)---(g). Also, by observing Figs.~\ref{Fig interference}(e)---(g) we find that the residual extension of the SSPN distribution is along the $y=x$ direction. This is in consistence with the sublattice strategy which cannot suppress the diagonal NNN coupling in Eq. (\ref{Eqn NNND}), leading to the photon leakage mainly along this direction. The FB localization of the steady states can be further quantified by the localization factor defined as the ratio of the SSPN populated in the pumping sites versus the SSPN populated in the pumping sites and their nearest-neighboring unit-cells. Such localization factor versus the d.c. mixing $\mathcal{T}_{BC}^\mathrm{dc}$ (and thus the flatness of the synthesized FB) is calculated and plotted in Fig. \ref{Fig interference}(h), where the difference between the RM pumpings and the single-site pumping can be clearly discriminated, indicating the survival of the proposed FB localization even in the presence of the unwanted NNN hopping channels.

The measurement of the proposed FB localization should also be considered. As the system is linear (i.e. it does not involve photon-photon interaction), the steady state can be described in the picture of multi-mode coherent state, and the SSPN on a particular site $(\mathbf{r},\alpha)$ can be measured by the method shown in Fig. \ref{Fig Lattice}: We capacitively connect this site to an external coil with an (optional for pumping) input and an output port. The steady state can be prepared by injecting microwave pulses through the input port for a sufficiently long time. During the steady-state period, energy will leak out from the coupling capacitance, which is proportional to the energy $\omega_\alpha \langle \alpha_{\mathbf{r}}^{\dagger} \alpha_{\mathbf{r}}\rangle$ with the proportional constant determined by the coupling capacitance \cite{MilburnQOBook}. The target observable $\langle \alpha_{\mathbf{r}}^{\dagger}\alpha_{\mathbf{r}}\rangle$ can then be measured by integrating the energy flowing to the output port in a given steady-state time duration, and the proposed FB localization can therefore be extracted by monitoring only few sites of the lattice (i.e. the pumping sites and their neighbors).  Actually, this scheme has already been used in a recent experiment, where both the amplitude and the relative phase of a TLR coherent state were measured \cite{NISTCoherentStateAPL2015}. The key point is that what we want to measure here is merely the expectation value $\langle \alpha_{\mathbf{r}}^{\dagger}\alpha_{\mathbf{r}}\rangle$ but not the detailed probability distribution in the TLRs Fock basis. It is this weak requirement that greatly simplify the measurement setup.

\section{Conclusion and Outlook}
\label{Sec Conclusion}
In conclusion, we have shown that it is not only possible but also advantageous to implement and detect the FB physics of Lieb lattice in the proposed circuit QED lattice. While the localized steady states of the RM excitations considered in this manuscript can be thoroughly understood in the single particle picture, what is more important is that the dispersionless flat band is an ideal platform of achieving correlated many-body states \cite{BergholtzFB2013,SondhiFB2013}. The introduction of interaction will lead us to the realm where rich but less explored physics locates. On the other hand, as the strong coupling between the TLRs and multi-level superconducting qubits has already been achieved \cite{DevoretReview2013,JQYouReview}, the Bose-Hubbard type \cite{RebicKerr2009PRL,HuYongCrossKerr2011} and Jaynes-Cummings-Hubbard type photon-photon interaction \cite{KochReview1,KochReview2} can be incorporated by coupling the TLRs with superconducting qubits. Therefore, our further direction should be the implementation and characterization of nonequilibrium photonic fractional Chern insulators in the proposed architecture \cite{MaciejkoNPFTI2015}.

\begin{acknowledgments}
We thank Z. D. Wang, M. Gong, and D. W. Zhang for helpful discussions. This work was supported in part by the National Fundamental Research Program of China (Grants No.~2012CB922103 and No.~2013CB921804), the National Science Foundation of China (Grants No.~11374117, No.~11375067, No. 11274129, and No. 11534001), and the PCSIRT (Grant No.~IRT1243).
\end{acknowledgments}

\appendix

\section{Eigenmodes of the lattice and their coupling}
\label{App Eigen}

In this Appendix, we analyze in detail the eigenmodes of the lattice and the coupling between them induced by the grounding SQUIDs. These two issues can be illustrated through the analysis of the highlighted unit-cell shown in Figs.~\ref{Fig Lattice}(a) and \ref{Fig Lattice}(c). During this investigation, we also estimate the parameters of the proposed circuit based on recently reported experimental data \cite{DCEexperimentNature2011,NISTParametricConversionNP2011,NISTHongOuMandelPRL2012,NISTParametricCouplingPRL2014,NISTCoherentStateAPL2015} and propose their representative values which have already been shown in Tab. \ref{Tab para}. As we focus solely on the highlighted unit-cell, the influence from the other part of the lattice are minimized by setting infinitesimal inductances for the grounding SQUIDs at the three individual ends.

\subsection{Eigenmodes of the unit-cell}

We assume the common grounding SQUID of the three TLRs has effective Josephson energy $E_{\mathrm{J}}=E_{\mathrm{J0}} \cos( \pi \Phi_{\mathrm{ext}} / \Phi_{\mathrm{0}})$ with $E_{\mathrm{J0}}$ its maximal Josephson energy, $\Phi_{\mathrm{ext}}$ the external flux bias, and $\Phi_{\mathrm{0}}=h/2e$ the flux quantum, as highlighted in Fig.~\ref{Fig Lattice}(a). In the first step let us assume that only a d.c. flux bias $\Phi_{\mathrm{ex}}^\mathrm{dc}$ is added. Physically speaking, a particular TLR (e.g. the TLR A) can hardly ``feel" the other two because the currents from them will flow mostly to the ground through the SQUID due to its very small inductance \cite{FelicettiPRL2014,WangYPChiral2015}. The SQUID can then be regarded as a low-voltage shortcut of the three TLRs, and it is this boundary condition that allows the definition of individual TLR modes in the coupled circuit. More explicitly, the Lagrangian of the unit-cell can be written as
\begin{align}
\mathcal{L}&= \sum_{\alpha} \int_{0}^{L_{\alpha}}\, \mathrm{d}x \, \frac{1}{2} [c(\frac{\partial \phi_{\alpha}(x,t)}{\partial t})^2-\frac{1}{l}(\frac{\partial \phi_{\alpha}(x,t)}{\partial x})^2]  \notag\\
&+\frac{1}{2}C_{\mathrm{J}}\dot{\phi}_{\mathrm{J}}^2+E_{\mathrm{J}}\cos (\frac{ \phi_{\mathrm {J}} }{\phi_{\mathrm{0}}}) \\
&\approx \sum_{\alpha} \int_{0}^{L_{\alpha}}\, \mathrm{d}x \, \frac{1}{2}[c(\frac{\partial \phi_{\alpha}(x,t)}{\partial t})^2-\frac{1}{l}(\frac{\partial \phi_{\alpha}(x,t)}{\partial x})^2]  \notag\\
&+\frac{1}{2}C_{\mathrm{J}} \dot{\phi}_{\mathrm{J}}^2-\frac{1}{2L_{\mathrm{J}}}\phi_{\mathrm{J}}^2
\label{eq:Lagrangian2}
\end{align}
with $c/l$ the capacitance/inductance per unit length of the TLRs, $\alpha=\mathrm{A}$, $\mathrm{B}$, $\mathrm{C}$ the label of the three TLRs, $L_{\alpha}$ the length of the $\alpha$th TLR,  $C_{\mathrm{J}}$ the capacitance of the SQUID, $\phi_{\mathrm{0}}=\Phi_{0}/2\pi$ the reduced flux quantum, $L_{\mathrm{J}}=\phi_{\mathrm{0}}^2/E_{\mathrm{J}}$ the effective inductance of the SQUID, $V_{\alpha}(x,t)$ the voltage distribution on the TLR $\alpha$, $\phi_{\alpha}(x,t)=\int_{-\infty}^{t} \mathrm{d}t' \,V_{\alpha}(x,t')$ the corresponding node flux distribution, $V_{\mathrm{J}}(t)$ the voltage across the grounding SQUID, and $\phi_{\mathrm{J}}(t)=\int_{-\infty}^{t} \mathrm{d}t' \, V_{\mathrm{J}}(t')$. In deriving Eq.~(\ref{eq:Lagrangian2}), we have linearized the grounding SQUID as $E_{\mathrm{J}}\cos(\phi_{\mathrm{J}}/\phi_{\mathrm{0}})\approx -\phi_\mathrm{J}^2/2L_\mathrm{J}$. This assumption is consistent with the described shortcut boundary condition and will be self-consistently verified later.

\begin{figure}[tbh!]
\centering
\includegraphics[width=0.48\textwidth]{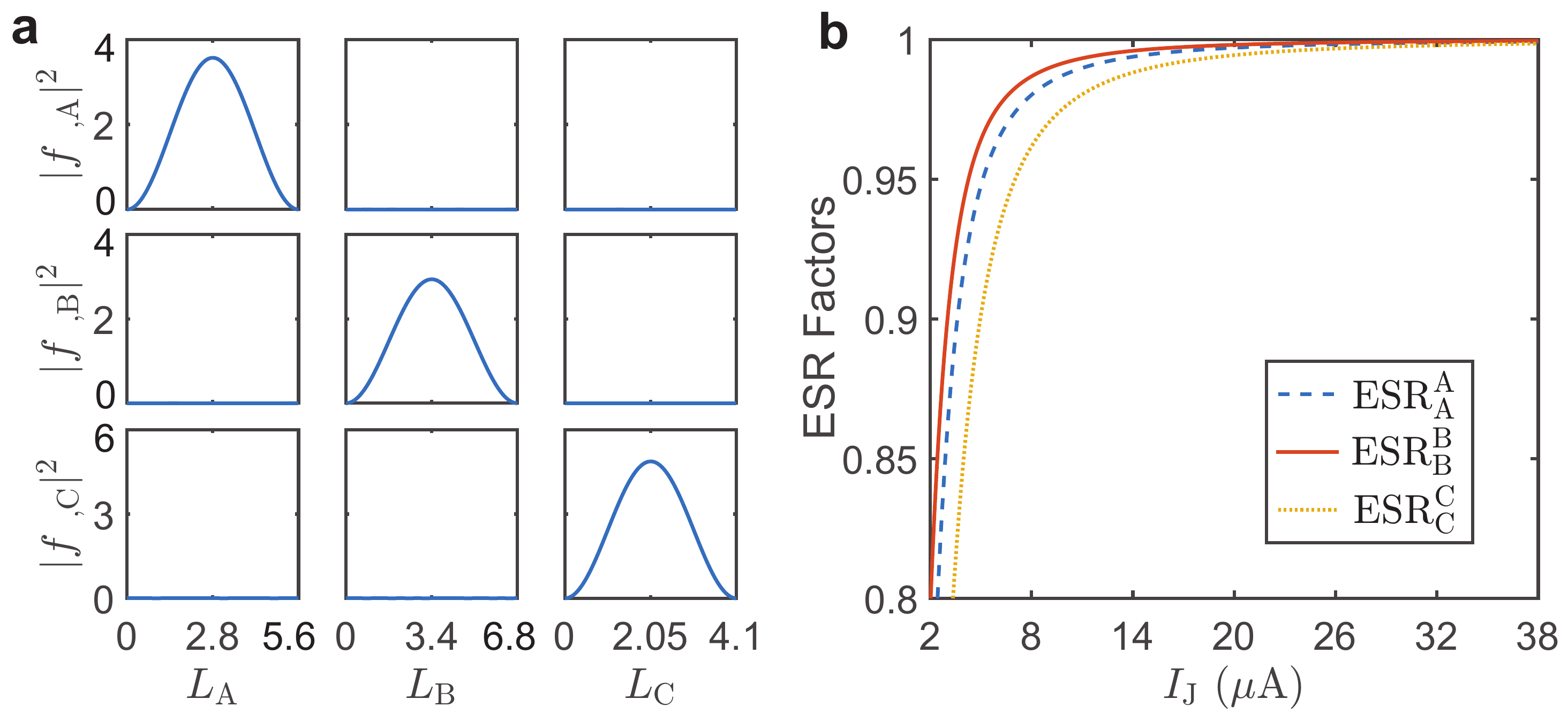}
\caption{\label{Fig Eigenmode}(Color Online) (a) Normalized node flux distributions of the lowest three eigenmodes in the highlighted unit-cell. $L_\alpha$ and $|f_{\alpha,n}|^2$ are in units of $\mathrm{mm}$ and $10^2 \mathrm{m^{-1}}$, respectively. (b) ESR factors of the three eigenmodes in their corresponding TLRs versus $I_{\mathrm{J}}$.}
\end{figure}

The equation of motion of $\phi_\alpha$ has the wave equation form
\begin{align}
\frac{\partial ^{2}\phi _{\alpha }}{\partial x^{2}}-\frac{1}{v^{2}}\frac{\partial^{2}\phi _{\alpha }}{\partial t^{2}}=0, \label{eq:wavequation}
\end{align}
with $v=1/\sqrt{cl}$, and the boundary conditions
\begin{eqnarray}
\phi_{\alpha}(x=0)=0,\,\phi_{\alpha}(x=L_{\alpha})=\phi_\mathrm{J}, \label{eq:boundaryconditon01}\\
-\frac{1}{l}\sum _{\alpha }\frac{\partial \phi_{\alpha} }{\partial x} |_{x=L_{\alpha}} =\frac{\phi _{\mathrm{J}}}{L_{\mathrm{J}}}+C_{\mathrm{J}}\ddot{\phi}_{\mathrm{J}}, \label{eq:boundaryconditon02}
\end{eqnarray}
can be obtained from Kirchhoff's law. The variable separation ansatz $\phi_{\alpha}(x,t)=\sum_{m}f_{\alpha,m}(x)g_m(t)$ is then exploited with $m=A,B,C$ the index of the eigenmodes. From Eq.~(\ref{eq:boundaryconditon01}) we have $f_{\alpha,m}(x)=C_{\alpha,m}\sin(k_m x)$, and by inserting $f_{\alpha,m}(x)$ into Eq.~(\ref{eq:boundaryconditon02}) we get
\begin{align}
& \sum _{\beta}C_{\beta,m }L_{\mathrm{J}}k_m\cos \left( k_mL_{\beta }\right) \notag \\
+{} &\left( l-\frac{C_{\mathrm{J}}L_{\mathrm{J}}}{c}k_m^{2}\right) C_{\alpha,m}\sin (k_mL_{\alpha})=0,\label{eq:transcendental}
\end{align}
which completely determine $f_{\alpha,m}(x)$ up to a normalization constant. Eq.~(\ref{eq:transcendental}) can be solved numerically with its typical solution plotted in Fig.~\ref{Fig Eigenmode}(a). Here we use the orthonormality relation \cite{LiebSQUIDCouplingNJP2012}
 \begin{align}
& \sum _{\beta }\int_0^{L_{\beta}}\mathrm{d}x\,f_{\beta ,m}(x)f_{\beta,n}(x) \notag \\
+{}& \frac{C_{J}}{c}f_{\alpha ,m}\left( L_{\alpha }\right) f_{\alpha,n}\left( L_{\alpha }\right) =\delta _{mn}.\label{eq:orthonormality}
\end{align}
and exploit the circuit parameters listed in Tab. \ref{Tab para}, which are chosen from recent experiments of dynamic Casimir effect and PFC in circuit QED \cite{DCEexperimentNature2011,NISTParametricConversionNP2011,NISTHongOuMandelPRL2012,NISTParametricCouplingPRL2014,NISTCoherentStateAPL2015}. Fig.~\ref{Fig Eigenmode}(a) demonstrates that the eigenmodes are well-separated in the corresponding TLRs, indicating the one-to-one correspondence between the TLRs and the eigenmodes. Such separation can be quantified by the energy storing ratio (ESR) factors of the $m$th mode in the $\alpha$th TLR, defined as
\begin{equation}
\mathrm{ESR}_m^\alpha=E_m^\alpha/E_m,
\end{equation}
with
\begin{align}
E_m^\alpha=\int_{0}^{L_{\alpha}}\, \mathrm{d}x \, \frac{1}{2}[c\omega_{m}^{2}+\frac{1}{l}{k}_{m}^{2}]f_{\alpha,m}^2(x),
\end{align}
\begin{align}
E_m &= \sum_{\alpha} \int_{0}^{L_{\alpha}}\, \mathrm{d}x \, \frac{1}{2}[c\omega_{m}^{2}+\frac{1}{l}{k}_{m}^{2}]f_{\alpha,m}^2(x)  \notag\\
&+\frac{1}{2}[C_{\mathrm{J}} \omega_{m}^{2} +\frac{1}{L_{\mathrm{J}}}]f_{\alpha,m}^2(x=L_\alpha),
\end{align}
and $\omega_m=vk_m$. For the $m$th mode, $\mathrm{ESR}_m^\alpha$ represents obviously the energy stored in the TLR $\alpha$ versus the whole energy of the mode. In Fig.~\ref{Fig Eigenmode}(b) three $\mathrm{ESR}_\alpha^\alpha$ factors versus varying $I_\mathrm{J}$ are calculated and shown with the other parameters leaved unchanged. The increase of the three $\mathrm{ESR}_\alpha^\alpha$ with increasing $I_\mathrm{J}$ can be noticed, and they are all above $0.99$ when $I_\mathrm{J}$ approaches the proposed $30$ $\mu\mathrm{A}$ in Tab. \ref{Tab para}, implying the well separation of the three eigenmodes.

The quantization of the eigenmodes is then straightforward. The Lagrangian $\mathcal{L}$ can be transformed to
\begin{align}
\mathcal{L}=\sum _{m}\frac{c\dot{g}_{m}^{2}}{2}-\frac{c\omega _{m}^{2}g_{m}^{2}}{2},
\label{eq:Lagrangian3}
\end{align}
with the help of Eq.~(\ref{eq:orthonormality}), and the corresponding Hamiltonian can be further be derived as
\begin{align}
\mathcal{H}_{\mathrm{0}}=\sum _{m}\frac{\pi _{m}^{2}}{2c}+\frac{c\omega _{m}^{2}g_{m}^{2}}{2},
\end{align}
with $\pi _{m}=\partial \mathcal{L}/\partial \dot{g}_{m}$ the canonical momentum of $g_{m}$. Through the definition of the creation/annihilation operators
\begin{align}
a_{m}^{\dag } =\sqrt{\frac{\omega_m c}{2\hbar }}g_{m}-i\sqrt{\frac{1}{%
2\hbar \omega_m c}}\pi _{m},\label{eq:creationoperator} \\
a_{m} =\sqrt{\frac{\omega_m c}{2\hbar }}g_{m}+i\sqrt{\frac{1}{2\hbar \omega_m c%
}}\pi _{m},\label{eq:annihilationopeartor}
\end{align}
$\mathcal{H}_0$ can finally be written as
\begin{align}
\mathcal{H}_{\mathrm{0}}=\sum _{m}\hbar \omega _{m}(a_{m}^{\dag }a_{m}+\frac{1}{2}),
\label{eq:eigenhamiltonian}
\end{align}
which reproduces exactly the one unit-cell version of Eq.~(\ref{Eqn Hami}).

\subsection{The grounding SQUID: d.c. mixing and linearization}

Here we temporarily stop to check the role played by the grounding SQUID. Firstly, $\phi_\mathrm{J}$ can be written as
\begin{equation}
\label{eq:phiJ1}
\phi_\mathrm{J}=\sum_{m} \phi^{m}(a_m+a_m^\dagger),
\end{equation}
with $\phi^{m}=f_{\alpha,m}(x=L_\alpha)\sqrt{\hbar/2\omega_m c}$ the r.m.s. node flux fluctuation of the $m$th mode across the grounding SQUID. With the parameters in Tab. \ref{Tab para} we have
\begin{align}
&(\phi^A,\phi^B,\phi^C)/\phi_0 \notag \\
={}&(1.6,1.9,3.1)\times 10^{-3}.
\end{align}
 Such small fluctuation of $\phi_\mathrm{J}$ indicates that the derived eigenmodes can be regarded as the individual $\lambda/2$ modes of the TLRs slightly mixed by the grounding SQUID with small but finite inductance (see also Fig.~\ref{Fig Eigenmode}). We then estimate to what extent the grounding SQUID mixes the individual $\lambda/2$ modes of the TLRs. We recall that such mixing can be physically traced back to the d.c. Josephson coupling
\begin{align}
\mathcal{E}_{\mathrm{dc}}&=-E_{\mathrm{J}}\cos \left( \frac{\phi_{\mathrm{J}}}{\phi_{\mathrm{0}}} \right)
\notag\\
&\approx \frac{1}{2} \left( \frac{\phi_{\mathrm{J}}}{\phi_{\mathrm{0}}} \right)^2 E_{\mathrm{J0}}\cos \left(\frac{\Phi_{\mathrm{ex}}^{\mathrm{dc}}}{2\phi _{0}}\right)
\notag\\
&=\sum _{m,n}\mathcal{T}_{mn}^{\mathrm{dc}}(a_{m}^{\dagger }+a_{m})(a_{n}^{\dagger }+a_{n}),
\label{eq:EJosillation}
\end{align}
with
\begin{eqnarray}
\mathcal{T}_{mn}^{\mathrm{dc}}=\frac{\phi^m\phi^n}{\phi_0^2}E_{\mathrm{J0}}\cos \left(\frac{\Phi_{\mathrm{ex}}^{\mathrm{dc}}}{2\phi _{0}}\right).
\end{eqnarray}
$\mathcal{T}_{mn}^{\mathrm{dc}}$ can then be regarded as the d.c. mixing between the individual $\lambda/2$ modes induced by the static bias of the grounding SQUID. Based on the parameters shown in Tab. \ref{Tab para}, we have the further estimation
\begin{equation}
\mathcal{T}_{mn}^{\mathrm{dc}}/2\pi \in [45,60] \text{ MHz} \approx \left[0.02,0.03\right] \Delta/2\pi ,
\end{equation}
which is in consistence with the previous presentation that the grounding SQUID only slightly mixes the original $\lambda/2$ modes of the TLRs.

We can also estimate the higher fourth order nonlinear term of $-E_{\mathrm{J}}\cos(\phi_{\mathrm{J}}/\phi_0)$ as
\begin{align}
    \mathcal{E}_{\mathrm{dc}}^4 &\approx \frac{1}{48} \left( \frac{\phi^j}{\phi_{\mathrm{0}}} \right)^4 E_{\mathrm{J0}}\cos \left(\frac{\Phi_{\mathrm{ex}}^{\mathrm{dc}}}{2\phi _{0}}\right) \notag\\
    &\in 2\pi \left[ 10^{-2},10^{-1} \right] \text{ kHz} \approx 10^{-6} \mathcal{T}^{\mathrm{dc}}_{mn},
\end{align}
i.e. six orders of magnitude smaller than the second-order terms reserved in Eqs.~(\ref{eq:Lagrangian2}) and (\ref{eq:EJosillation}). Such small term can be safely neglected and the validity of the Taylor expansion in deriving Eq.~(\ref{eq:Lagrangian2}) is therefore verified in a self-consistent way.

\subsection{Parametric coupling between the eigenmodes}

The parametric coupling between the three eigenmodes originates from the dependence of $E_{\mathrm{J}}$ on $\Phi_{\mathrm{ext}}$
\begin{align}
E_{\mathrm{J}}&=E_{\mathrm{J0}}\cos \left[\frac{1}{2\phi _{0}}\left(\Phi_{\mathrm{ex}}^{\mathrm{dc}}+\Phi_{\mathrm{ex}}^{\mathrm{ac}}(t)\right)\right]   \\
&\approx E_{\mathrm{J0}}\cos \left(\frac{\Phi_{\mathrm{ex}}^{\mathrm{dc}}}{2\phi _{0}}\right)-\frac{E_{\mathrm{J0}}\Phi_{\mathrm{ex}}^{\mathrm{ac}}(t)}{2\phi _{0}}\sin \left(\frac{\Phi_{\mathrm{ex}}^{\mathrm{dc}}}{2\phi _{0}}\right),
\label{eq:EJosillation2}
\end{align}
where we have assumed that a small a.c. fraction $\Phi_{\mathrm{ex}}^{\mathrm{ac}}(t)$ has been added to $\Phi_{\mathrm{ext}}$ with $\left|\Phi_{\mathrm{ex}}^{\mathrm{ac}}(t)\right| \ll \left|\Phi_{\mathrm{ex}}^{\mathrm{dc}} \right|$.
 %such that the first order expansion at the d.c. bias point $\Phi_{\mathrm{ex}}^{\mathrm{dc}}$ can be performed.
As stated in the previous main text, $\Phi_{\mathrm{ex}}^{\mathrm{ac}}(t)$ is composed of two tones \begin{align}
\label{Eqn twotone}
\Phi_{\mathrm{ex}}^{\mathrm{ac}}(t)&=\Phi_{\mathrm{CA}}\cos ( 2 \Delta t - \theta_{CA} )+\Phi_{\mathrm{BA}}\cos ( \Delta t + \theta_{\mathrm{BA}} )
\end{align}
where the $2\Delta$ tone is exploited to induce the vertical $\mathrm{A} \Leftrightarrow \mathrm{C}$ hopping, and the $\Delta$ tone is used for the horizontal $\mathrm{A} \Leftrightarrow \mathrm{B}$ hopping \cite{WangYPChiral2015}. By representing $\phi_{\mathrm{J}}$ as the form shown in Eq.~(\ref{eq:phiJ1}) we obtain the a.c. coupling from the second term of Eq.~ (\ref{eq:EJosillation2})
\begin{equation}
\mathcal{H}_{\mathrm{a.c.}}=\frac{E_{\mathrm{J0}}\Phi_{\mathrm{ex}}^{\mathrm{ac}}(t)}{4\phi _{0}^3}\sin \left(\frac{\Phi_{\mathrm{ex}}^{\mathrm{dc}}}{2\phi _{0}}\right)
\left[\sum _{m} \phi^m \left(a_m+a_m^{\dagger}\right)\right]^2,
\label{eq:HDCcoupling2}
\end{equation}
In the rotating frame of $\mathcal{H}_{\mathrm{S}}$, the induced parametric photon hopping between the TLRs can be further written as
\begin{align}
\label{Eqn effHami}
\mathcal{H}_{\mathrm{L}}&=e^{it\mathcal{H}_{\mathrm{S}}} \mathcal{H}_{\mathrm{a.c.}} e^{-it\mathcal{H}_{\mathrm{S}}} \notag\\
& \simeq \left[\mathcal{T}_\mathrm{BA}e^{i\theta_\mathrm{BA}}B^{\dagger}A
+\mathcal{T}_\mathrm{CA}e^{i\theta_\mathrm{CA}}C^{\dagger}A \right]+\mathrm{h.c.},
\end{align}
where $\mathcal{T}_{\alpha\beta}$ are the effective hopping strengths proportional to the corresponding $\Phi_{\alpha\beta}$ in Eq. (\ref{Eqn twotone}), and the fast-oscillating terms in $e^{it\mathcal{H}_{\mathrm{S}}} \mathcal{H}_{\mathrm{AC}} e^{-it\mathcal{H}_{\mathrm{S}}}$ are omitted due to rotating wave approximation. The amplitudes of the two tones can be selected as $\left[ \Phi_{\mathrm{BA}}, \Phi_{\mathrm{CA}}\right]=\Phi_0\left[ 0.9\%, 1.3\% \right]$ such that the homogeneous coupling strength $\mathcal{T}/2\pi=\mathcal{T}_{\mathrm{BA}}/2\pi=\mathcal{T}_{\mathrm{CA}}/2\pi=10\,\mathrm{MHz}$ can be induced \cite{NISTParametricConversionNP2011,NISTHongOuMandelPRL2012,NISTParametricCouplingPRL2014,NISTCoherentStateAPL2015}. In addition, Eqs. (\ref{Eqn twotone}) and (\ref{Eqn effHami}) imply that arbitrary nontrivial hopping phases can be obtained by the appropriate choice of the initial phases of the modulating pulses, indicating the site-resolved synthesization of the artificial gauge field for the microwave photons. For instance, we can construct the nontrivial horizontal $\mathrm{A}\Leftrightarrow\mathrm{B}$ hopping phases while leave the vertical $\mathrm{A}\Leftrightarrow\mathrm{C}$ hopping phases trivial. Such configuration leads to Landau gauge
\begin{align}
  \mathbf{A}&=\left[A_x,0,0 \right], \notag\\
  \mathbf{B}&=B\mathbf{e}_z=\left[ 0,0,-{\partial A_x}/{\partial y} \right],
\end{align}
which will be exploited in the main text.

We should also be careful that the modulating frequency of $\Phi^{\mathrm{ac}}_{\mathrm{ex}}(t)$ must be lower than the plasma frequency of the grounding SQUID $\omega_{\mathrm{p}}=\sqrt{8E_{\mathrm{C}} E_\mathrm{J}}$ \cite{KochTransmonPRA2007}, otherwise the internal degrees of freedom of the SQUID will be activated and complex quasi-particle excitations will emerge \cite{FelicettiPRL2014}. This requirement is fulfilled by the very small inductance of the grounding SQUID. With the parameters selected we have the estimation $\omega_{\mathrm{p}}/2\pi \approx 136\,\mathrm{GHz}=68\Delta/2\pi$, leading to the effective suppression of the grounding SQUID excitation.

\section{Low frequency noise of the lattice}
\label{App noise}
In this Appendix, we calculate in detail the fluctuation induced by the flux and critical current $1/f$ noises. It is generally believed that a particular fluctuation $\delta O(t)$ of the physical variable $O$ in solid-state physics exhibiting the $1/f$ spectrum can be modelled by the Dutta-Horn model, i.e. the summation of random telegraph noises emitted from an ensemble of bistable fluctuators \cite{FlickerRMP2014}. The $1/f$ type fluctuation of $\delta O$ can be described by its noise spectrum
\begin{align}
S_{O} (\omega) &=\int_{-\infty}^{+\infty} \, \mathrm{d}t e^{i\omega t} \langle \delta O(t) \delta O(0) \rangle \notag\\
&=\frac{2\pi \mathcal{A}_O^2}{\omega},\,\omega \in \left[ \omega_{\mathrm{min}}, \omega_{\mathrm{max}} \right],
\end{align}
where $\mathcal{A}_O$ labels the noise spectrum at $2\pi \times 1\,\mathrm{Hz}$, taking the same dimension of $\delta O$, and $\omega_{\mathrm{min}}/\omega_{\mathrm{max}}$ denote the lower/upper cutoff of the $1/f$ spectrum, respectively. In the following calculation, we set
\begin{align}
\omega_{\mathrm{min}}/2\pi=1\,\text{Hz}, \omega_{\mathrm{max}}/2\pi=1\,\text{GHz},
\end{align}
based on the scale of the experiment time and the $\sim 50$ $\mathrm{mK}$ temperature scale of the dilute refrigerator \cite{IthierDecoherencePRB2005,FlickerRMP2014}. In addition, we can treat $\delta O(t)$ as quasi-static in the following estimation due to its low frequency property, i.e. it does not vary during a experimental run, but varies between different runs. The variance of $\delta O(t)$ can be evaluated from $S_O(\omega)$ as
\begin{align}
\langle \left(\delta O (t) \right) ^2 \rangle
&=\frac{1}{2\pi}\int \, \mathrm{d}\omega \int_{-\infty}^{+\infty} \, \mathrm{d}t e^{i\omega t} \langle \delta O(t) \delta O(0) \rangle    \notag\\
&=\frac{1}{2\pi}\int \, \mathrm{d}\omega S_O\left( \omega \right)\approx \mathcal{A}^2_O \left(\ln \gamma_{\mathrm{max}}-\ln \gamma_{\mathrm{min}} \right),
\end{align}
indicating that the range of the fluctuating $\delta O$ can be roughly estimated as $\delta O \in \left[-5,5\right]\mathcal{A}_O$.

In the following we estimate the influence of the $1/f$ noises on the proposed scheme. For the flux type $1/f$ noise, various previous measurements has shown that $\mathcal{A}_\Phi/\Phi_0 \in  \left[10^{-6},10^{-5}\right]$ does not vary greatly with the loop size, inductor value, or temperature \cite{FluxqubitFlickerPRL2006,MartinisFlickerPRL2007,FluxQubit1fGeometryPRB2009}. Therefore the strength of $\delta \Phi$ can be estimated as $\delta \Phi/\Phi_0 \in [10^{-5},10^{-4}]$. Such fluctuation is by two orders of magnitude smaller than the d.c. $\Phi_\mathrm{ex}^\mathrm{dc}=0.37 \Phi_0$ and the a.c. $[\Phi_\mathrm{BA},\Phi_\mathrm{CA}]=\left[ 0.9\%, 1.3\% \right]\Phi_0$. The existence of $\delta \Phi$ shifts $\Phi_\mathrm{ex}^\mathrm{dc}$ in a quasi-static way, and its influence can be evaluated through the Taylor expansion of Eqs. (\ref{eq:EJosillation}) and (\ref{eq:HDCcoupling2}) with respect to $\Phi_\mathrm{ex}^\mathrm{dc}$:
\begin{align}
\delta \mathcal{E}_{\mathrm{dc}} &\approx \frac{\delta \Phi}{4 \phi_0^3} E_{\mathrm{J0}} \sin \left(\frac{\Phi_{\mathrm{ex}}^{\mathrm{dc}}}{2\phi _{0}}\right)
\left[ \sum_m \, \phi^m \left( a_m+a_m^\dagger \right) \right]^2,
\label{eq:fluxdiagonal}
\end{align}
\begin{align}
\delta\mathcal{H}_{\mathrm{a.c.}}&=\frac{E_{\mathrm{J0}}\Phi_{\mathrm{ex}}^{\mathrm{ac}}(t)\delta\Phi}{8\phi _{0}^4}\cos \left(\frac{\Phi_{\mathrm{ex}}^{\mathrm{dc}}}{2\phi _{0}}\right)
\left[\sum _{m} \phi^m\left(a_m+a_m^{\dagger}\right)\right]^2.
\label{eq:fluxoffdiagonal}
\end{align}
Based on the parameters in Tab. \ref{Tab para}, we can evaluate that the fluctuating $\delta \Phi$ causes negligible
\begin{align}
\delta \omega_{\mathbf{r},\alpha}/2\pi &\in [10^{-3},10^{-2}] \text{ MHz} < 10^{-3} \mathcal{T}/2\pi, \\
\delta \mathcal{T}_{\mathbf{r},\alpha}^{\mathbf{r^\prime},\beta}/2\pi &\in [10^{-4},10^{-3}] \text{ MHz} < 10^{-4} \mathcal{T}/2\pi .
\end{align}

In addition, experiments have shown that the critical current noise has $\mathcal{A}_{I_\mathrm{J0}} \approx 10^{-6} I_{\mathrm{J0}}$ for a junction at temperature $4\,\mathrm{K}$ \cite{JohnClarkICPRB2004,MartinisFlickerPRL2007}. The parameter $\mathcal{A}_{I_\mathrm{J0}}/I_\mathrm{J0}$ proves to be proportional to the temperature down to at least $100\,\mathrm{mK}$. Therefore we set $\mathcal{A}_{I_\mathrm{J0}}/I_\mathrm{J0} \in [10^{-7},10^{-6}]$. The influence of the critical current noise can also be estimated by the Taylor expansion of $\mathcal{E}_\mathrm{dc}$ and $\mathcal{H}_\mathrm{a.c.}$ with an alternative respect to $E_\mathrm{J0}=I_\mathrm{J0}\hbar/2e$. Following the estimation similar to that of the previous flux noise, we can evaluate that the fluctuating $\delta I_{\mathrm{J0}}$ causes
\begin{align}
\delta \omega_{\mathbf{r},\alpha}/2\pi &\in [10^{-4},10^{-3}] \text{ MHz} < 10^{-4} \mathcal{T}/2\pi, \\
\delta \mathcal{T}_{\mathbf{r},\alpha}^{\mathbf{r^\prime},\beta}/2\pi &\in [10^{-5},10^{-4}] \text{ MHz} < 10^{-5} \mathcal{T}/2\pi ,
\end{align}
which are even smaller than the effects induced by the flux noises.

\section{Band structure calculation of the Lieb lattice}
\label{App Band}
The energy band structure shown in Figs. \ref{Fig Lattice}(c) and \ref{Fig Lattice}(d) are calculated by using the following general procedure \cite{BergholtzFB2013,BernevigBook} with periodic boundary condition imposed: Consider a quadratic Hamiltonian
\begin{equation}
\mathcal{H}_0=\sum_{n,m} t^{ab}_{nm}c^\dagger_{n,a}c_{m,b} \ ,
\end{equation}
where $n,m=1,\ldots,N_s$ label the $N_s$ unit-cell on the lattice, $a,b$ label the orbits inside the unit-cell, and $t^{ab}_{nm}=t^{ab}_{n-m}$ is the translation-invariant coupling strength. By using $c^\dagger_{{\mathbf{k}},a}={N_s}^{-1/2}\sum_n e^{i{\mathbf{k}}\cdot {\mathbf{R}}_n}c^\dagger_{n,a}$, we transform $\mathcal{H}_0$ to
\begin{equation}
H_0=\sum_{a,b,{\mathbf{k}}} \mathcal H_\mathbf{k}^{ab}c^\dagger_{{\mathbf{k}},a}c_{{\mathbf{k}},b}
\ ,
\end{equation}
where ${\mathbf{k}}=(k_x,k_y)$ is the single-particle pseudo-momentum in the first Brillioun zone, $\mathbf{R}_n$ is the location of the $n$th unit-cell, and
\begin{align}
\mathcal H_\mathbf{k}^{ab}&\equiv\frac 1 {N_s}\sum_{n,m} t^{ab}_{nm}e^{-i{\mathbf{k}}\cdot ({\mathbf{R}}_n-{\mathbf{R}}_m)} \notag\\
&=\sum_{n} t^{ab}_{n1}e^{-i{\mathbf{R}}\cdot ({\mathbf{R}}_n-{\mathbf{R}}_1)} \ .
\end{align}
The energy $E_s(\mathbf{k})$ for the orbit $s$ and the momentum $\mathbf{k}$ is then obtained by diagonalizing $\mathcal{H}_{\mathbf{k}}$ for each $\mathbf{k}$ separately. In particular, Fig. \ref{Fig Lattice}(c) and Eq. (\ref{Eqn flateigen}) are calculated for an ideal Lieb lattice with uniform nearest-neighbor hopping strength and zero magnetic field, i.e.
\begin{equation}
\mathcal{H}_{\mathbf{k}}=\mathcal{T}
\begin{bmatrix}
0&1+e^{ik_x/2}&1+e^{-ik_y/2}\\
1+e^{-ik_x/2}&0&0\\
1+e^{ik_y/2}&0&0
\end{bmatrix}
\ .
\end{equation}
Meanwhile, Fig. \ref{Fig Lattice}(d) is calculated when the NNN coupling in Eq. (\ref{Eqn NNND}) is taken into account, i.e. by diagonalizing $\mathcal{H}_{\mathbf{k}}$ with an additional NNN contribution
\begin{equation}
\mathcal{H}_{\mathbf{k}}^{\mathrm{NNN}}=\frac{2{\mathcal{T}_{\mathrm{BC}}^{\mathrm{dc}}}^2}{3\Delta}
\begin{bmatrix}
0&0&0\\
0&-\cos(k_x-k_y)&0\\
0&0&\cos(k_x-k_y)
\end{bmatrix}
\ ,
\end{equation}
added.

%\bibliography{TP}


%\section*{Author Contributions}
%Y.H. and J. H. Gao proposed the idea. Y.P.W. and Z.H.Y carried out all calculations under the guidance of Y.H., Y.W. participated in the discussions. Y.H and J.H.G contributed to the interpretation of the work and the writing of the manuscript.

%\section*{Competing Interests}
%The authors declare that they have no competing financial interests.

\end{document}